\documentclass[namedreferences]{solarphysics}
\usepackage[hyperref,optionalrh,solaromanenum]{spr-sola-addons} % For Solar Physics 
\usepackage{graphicx}                    % For eps figures, newer & more powerfull
\usepackage{color}                       % For color text: \color command
\usepackage{breakurl}                         % For breaking URLs easily trough lines
\usepackage{makecell}
\usepackage{siunitx}
\DeclareSIUnit{\arcsecond}{^{\prime\prime}}
\DeclareSIUnit{\sqrarcsecond}{arcsec\squared}

\begin{document}

\begin{article}

\begin{opening}

\title{Probing Upflowing Regions in the Quiet Sun and Coronal Holes}

%%%%%%%%%%%%%%%%%%%%%%%%%%%%%%%%%%%%%%%%%%%%%%%%%%%
%% Authors Names
%
% \author[addressref={},corref,email={}]{\inits{}\fnm{}\lnm{}\orcid{}}
\author[addressref={aff1,aff2},corref,email={conradsc@phys.ethz.ch}]{\inits{C.}\fnm{Conrad}~\lnm{Schwanitz}}
\author[addressref={aff2,aff1},corref,email={louise.harra@pmodwrc.ch}]{\inits{L.}\fnm{Louise}~\lnm{Harra}}
\author[addressref=aff3,corref,email={nour.raouafi@jhuapl.edu}]{\inits{N. E.}\fnm{Nour E.}~\lnm{Raouafi}}
\author[addressref=aff4,corref,email={alphonse.sterling@nasa.gov}]{\inits{A.C.}\fnm{Alphonse C.}~\lnm{Sterling}}
\author[addressref=aff5,corref,email={amoreno@iaa.es}]{\inits{A.}\fnm{Alejandro}~\lnm{Moreno Vacas}}
\author[addressref=aff5,corref,email={jti@iaa.es}]{\inits{J.C.}\fnm{Jose Carlos}~\lnm{del Toro Iniesta}}
\author[addressref=aff5,corref,email={orozco@iaa.es}]{\inits{D.}\fnm{David}~\lnm{Orozco Suárez}}
\author[addressref=aff6,corref,email={hirohisa.hara@nao.ac.jp}]{\inits{H.}\fnm{Hirohisa}~\lnm{Hara}}

%%%%%%%%%%%%%%%%%%%%%%%%%%%%%%%%%%%%%%%%%%%%%%%%%%%
%% Runningheads
%
\runningauthor{C. Schwanitz et al.}
\runningtitle{Probing coronal upflows}

%%%%%%%%%%%%%%%%%%%%%%%%%%%%%%%%%%%%%%%%%%%%%%%%%%%
%% Affilations 
%% id shold be the same with \author addressref value.
\address[id=aff1]{ETH Zürich, Institute for Particle Physics and Astrophysics, 8092 Zürich, Switzerland}
\address[id=aff2]{Physikalisch Meteorologisches Observatorium Davos, World Radiation Center, 7260 Davos, Switzerland}
\address[id=aff3]{ Johns Hopkins University Applied Physics Laboratory, Laurel, MD 20723-6099, USA}
\address[id=aff4]{NASA Marshall Space Flight Center, Huntsville, AL 35812, USA}
\address[id=aff5]{Instituto de Astrofísica de Andalucía, CSIC, 18008 Granada, Spain}
\address[id=aff6]{National Astronomical Observatory of Japan, Mitaka, Tokyo 181-8588, Japan}

%\address[id={}]{}

%%%%%%%%%%%%%%%%%%%%%%%%%%%%%%%%%%%%%%%%%%%%%%%%%%%
%%% Abstract 
\begin{abstract}
Recent observations from Parker Solar Probe have revealed that the solar wind has a highly variable structure. How this complex behaviour is formed in the solar corona is not yet known, since it requires omnipresent fluctuations, which constantly emit material to feed the wind. In this article we analysed 14 upflow regions in the solar corona to find potential sources for plasma flow. The upflow regions were derived from spectroscopic data from the EUV Imaging Spectrometer (EIS) onboard Hinode determining their Doppler velocity and defining regions which have blueshifts stronger than \SI[per-mode=symbol]{-6}{\kilo\metre\second\tothe{-1}}. To identify the sources of this blueshift data from the Atmospheric Imaging Assembly (AIA) and the Helioseismic and Magnetic Imager (HMI), onboard the Solar Dynamics Observatory (SDO), and the X-ray Telescope (XRT), onboard Hinode, were used. The analysis revealed that only 5 out of 14 of the upflows were associated with frequent transients, like obvious jets or bright points. In contrast to that, seven events were associated with small-scale features, which show a large variety of dynamics. Some resemble small bright points, while others show an eruptive nature, all of which are faint and only live for a few minutes; we can not rule out that several of these sources may be fainter and, hence, less obvious jets. Since the complex structure of the solar wind is known, this suggests that new sources have to be considered or better methods used to analyse the known sources. This work shows that small and frequent features, which were previously neglected, can cause strong upflows in the solar corona. These results emphasise the importance of the first observations from the Extreme-Ultraviolet Imager (EUI) onboard Solar Orbiter, which revealed complex small-scale coronal structures.
\end{abstract}

%%%%%%%%%%%%%%%%%%%%%%%%%%%%%%%%%%%%%%%%%%%%%%%%%%%
%% Keywords
%

\end{opening}
%-------------------------------------------------

%%%%%%%%%%%%%%%%%%%%%%%%%%%%%%%%%%%%%%%%%%%%%%%%%%%
%% Sections
%
% \section{}%\label{s:?} 

\section{Introduction}
\label{s:Intro}

The solar corona is transient with a wide range of phenomena from small to large scales.  Key questions regarding the solar corona remain unanswered. These include how the corona is heated and how the solar wind is formed. The solar wind is highly variable as can be seen clearly with the new observations from the Parker Solar Probe mission from a vantage point of 35 solar radii during the first two perihelia, e.g.  \cite{bourouaine2020turbulence}. There are many potential sources of the solar wind, and studies of frequent and small-scale phenomena show promise for feeding energy into the corona and solar wind. The first commissioning datasets from the EUV Imagers onboard Solar Orbiter indicate brightenings at the smallest scales so far seen in the corona from a vantage point of around 0.55 A.U. \citep{berghmans2021extreme}. These brightenings were found to have a power-law distribution, with the smallest brightenings seen to have 2 pixels (with an area of \SI{0.08}{\mega\metre\squared}).

Coronal holes are dark regions in the solar corona when viewed in X-rays. They are associated with `open' magnetic fields  \citep[for example, ][]{fisk2001behavior} that extend far into the heliosphere. They are of importance due to the fact that the fast solar wind originates from them. Coronal holes are also very dynamic with many phenomena being observed in the coronal emission lines. Both in coronal holes and the quiet Sun, a wide range of transient features exist, including plumes, jets, and bright points.

Coronal plumes are mostly seen in coronal holes. They are long and narrow, hazy structures, which extend into the heliosphere for several solar radii. They are best observed in white light and extreme ultraviolet. In contrast to other coronal transients, plumes can have relatively long lifetimes of up to a few days. Long-lived plumes are reported to disappear and reappear at the same location over two weeks. However, plumes also show changes within less than ten minutes. They are even connected to coronal jets and other transients, but the details of this connection are not fully understood yet \citep{raouafi2014role}.

Solar coronal jets \citep[for example, ][]{raouafi2016solar} are frequent and diverse transients, which might play a crucial role for the solar wind. Since their detection, they have been observed on different size scales from large jets at the edges of active regions to small ones seen in coronal holes. Their lifetime is usually 3-\SI{30}{\minute} \citep[e.g. ][]{nistico2009characteristics, kim2007small} with propagation speeds of 100-\SI[per-mode=symbol]{400}{\kilo\metre\second\tothe{-1}}. Coronal jets are observed in a large variety of morphologies. The most prominent types are standard jets and blowout jets \citep{moore2010dichotomy}. Models for standard jets were first proposed by \cite{shibata1992observations} and are based on reconnection in magnetic loops. This leads to a jet along the new open field line and a brightening at the base. The morphology of standard jets is usually very plain with a narrow shape. In contrast to that, the initial loop structure of blowout jets shows a strong interwinding. This leads to a cascade-like reconnection, which is the cause of a delayed blowout. Both of those well-described jet types are characterised by reconnecction, which leads to a well visible beam-like structure. Later on, we refer to both types by "classic jet" which is a feature that is easily recognisable as a coronal jet in AIA or XRT images. Such a designation is necessary because, as we shall see later, we find some sources to have some characteristics of jets but are much fainter and harder to make out than jets hitherto discussed in most jet papers. We do not know whether these features are actual jets and, therefore, do not want to use that term with them. We use the term "classic jet" here because the term "standard jet" would be confused with the standard/blowout-jet terminology introduced by \cite{moore2010dichotomy}.  A recent interpretation of jets involves mini-filament eruptions  \citep{sterling2015small, sterling2016minifilament}. Many coronal bright points were observed to have mini-filament eruptions \citep{hong2014coronal}. Some mini-filament eruptions then caused blow-out jets \citep{hong2011micro}. Jets are observed in many wavelengths. However, not all X-ray jets have a counterpart in EUV or vice versa \citep{raouafi2008evidence, moore2010dichotomy}. A consensus for the driving mechanisms of jets has not yet been reached. The two most discussed ideas are either based on flux emergence or flux cancellation from the photosphere \citep[for example, ][]{raouafi2016solar} beneath or on a local loss of stability, which leads to magnetic reconnection \citep[for example, ][]{raouafi2016solar}. Jets appear in a large variety and some phenomena might be related to each other and other coronal transients.
Previous studies have made use of Hinode/EIS to investigate the properties of both classical jet types. Spectroscopic rasters allowed to measure line-of-sight velocities up to \SI[per-mode=symbol]{279}{\kilo\metre\second\tothe{-1}} \citep{madjarska2011dynamics} and revealed twisting motions within the jet \citep{young2014coronal}. The length of the jet reaches up to \SI[per-mode=symbol]{87}{\mega\metre} from the bright point \citep{young2014coronal}.

Coronal bright points are important and frequent features in the solar corona. They are rooted in magnetic footpoints of opposite polarity and are visible as loops in EUV and X-rays. They are smaller than active regions but show many similarities to their larger active region counterparts. Their lifetimes are generally less than \SI{20}{\hour} \citep{golub1975observation, harvey1993lifetimes, alipour2015statistical}. Coronal bright points can be observed in the quiet Sun, coronal holes, and next to active regions. During their lifetime they can be the source of jets and filament eruptions, which possibly result from a restructuring and cancellation of the associated magnetic structure \citep{hong2014coronal}.

All of the presented transients have a local influence on the solar corona by ejecting plasma. The primary motivation for this article is to understand what the relevant sources of upflowing plasma in the quiet Sun and coronal holes are and how they may contribute to the solar wind. Jets have been observed many times and are a source of plasma upflow. More recently, a new feature, called dark jets, was first detected in solar coronal holes in a \SI{44}{\hour} long study of Hinode/EIS rasters \citep{young2015dark}, where the dark jets showed blueshifts in the line-of-sight velocity maps of the Fe XII 192.12 \AA\ emission line. However, there were no classic jets present in the corresponding SDO/AIA 193 \AA\ filter band. These dark jets were mostly present next to bright points either in the coronal hole or at the coronal hole boundary. The blueshifted regions show a large variety of shapes from elongated to fan-like structures. Their lifetime could not be determined precisely due to the raster time, but only narrowed down to the raster time of 62 min. None of the blueshifted regions reappear in successive rasters. However with 11 dark jets in the observation period of \SI{44}{\hour} they appeared nearly as frequent as classic jets of which 13 were observed.

The focus of this article is to probe upflowing plasma in coronal holes and the quiet Sun in EUV spectroscopic data. Using the spectroscopic data as a starting point means there is no requirement for a jet or an eruptive feature to be strong in intensity. This avoids a bias towards classic jets and allows the exploration of all possible mechanisms for the creation of upflowing plasma. To do this the line-of-sight Doppler velocities are derived from Hinode/EIS rasters and regions of interest are defined above a certain Doppler velocity level. By combining them with data from SDO/AIA and Hinode/XRT the origins of those sources are derived. SDO/HMI data could only be used for four events in more detail since most events are too close to the limb.

The next section describes the instrumentation and data analysis methods used, including the alignment of different data sets (Hinode/EIS, SDO/AIA, and Hinode/XRT). Section 3 presents the results of the study with a general overview of all the events followed by examples of four of the most common causes and an analysis of the HMI events. The article concludes with a summary and a discussion of the potential importance of these events for the solar wind.

\section{Data Analysis}
\label{s:Methods} 

The Hinode mission \citep{kosugi2007hinode} was launched in September 2006 to a sun-synchronous orbit. There are three instruments onboard: the Solar Optical Telescope (SOT) \citep{tsuneta2008solar}, the X-ray Telescope (XRT) \citep{golub2008x}, and the Extreme-Ultraviolet Imaging Spectrometer (EIS) \citep{culhane2007euv}. The locations of the blueshifted plasma were determined from the EIS data, while XRT is analysed to determine whether there are associated X-ray jets. EIS is an EUV spectrometer that has two wavelength bands 170\,--\,210$\,$\AA\ and 250\,--\,290$\,$\AA. Those emission lines cover temperatures from \SI{50}{\kilo\kelvin} to \SI{20}{\mega\kelvin}. The spatial resolution is about 2 arcsec over a maximum field of view of 560 $\times$ 512 arcsec$^2$.

The Hinode/EIS data for this work consists of 12 rasters taken from three different coronal hole studies. Hinode/EIS studies for coronal holes are characterised by long exposure times to improve the signal-to-noise ratio. Table \ref{table:events} summarises the times and durations of the rasters used for this project. The campaigns on 04 Feb. 2020 and 07 Mar. 2020 both used the \textit{HOP81\_new\_study} study with number 582. Each of their rasters has 81 pointings with a slit of \SI{2}{\arcsecond} and an exposure time of \SI{50000}{\milli\second}. The scan then has a stepsize of \SI{4}{\arcsecond}. The third campaign, which took place from 08 Feb. 2020 until 12 Feb. 2020 used the \textit{GDZ\_PLUME1\_2\_300\_150} study with number 537. It is very similar to the first one and only differs in the number of pointings, which are 75 here, and the exposure times, which are three times longer with \SI{350000}{\milli\second}.

\begin{table}
\begin{tabular}{ c|c|c|c } 
Date & \makecell{Number of \\ EIS rasters} & Raster Time & Event numbers\\ 
07 Mar. 2020 & 5 & \SI{1}{\hour}\SI{10}{\minute} & 1 - 5 $\;$ \\ 
04 Feb. 2020 & 3 & \SI{1}{\hour}\SI{10}{\minute} & 6 - 10 \\ 
08\,--\,12 Feb. 2020 & 4 & \SI{1}{\hour}\SI{10}{\minute} - \SI{3}{\hour}\SI{10}{\minute} & 11 - 14
\caption{Three different Hinode/EIS studies are used from February and March 2020. The duration time for most rasters is \SI{1}{\hour} and \SI{10}{\minute}. In each study a number of blueshift events are found, which are assigned a number to identify them.}
\label{table:events}
\end{tabular}
\end{table}

The Hinode/EIS rasters are processed with the SolarSoft calibration routine eis\_prep. The data are corrected for orbital variations using the eis\_wave\_corr routine. This analysis focuses on the Fe XII emission line at 195.12$\,$\AA.  The emission line is fitted by using eis\_auto\_fit routine, using a single Gaussian fit to determine intensity and Doppler shift velocity. There is a self-blend in the Fe XII line with 195.18$\,$\AA\ for which we tested a double Gaussian fit, but in these examples it made no significant impact on the Doppler velocities and the structures of the upflow features, so we used a single Gaussian fit. No significant blue asymmetries were seen in the spectra.  The correctness of the fits was checked by visual inspection for random samples from each raster, as well as checking the $\chi^2$of the fit. The fits are corrected for orbital variations with eis\_update\_fitdata, while the rest wavelength is determined from the average centroid of the raster. The Doppler-shift maps are then used to find the blueshifted features. The velocity maps are smoothed to allow for contours that are not disrupted by a single bad pixel. The smoothing consists of two steps: in a first step, regions in the Doppler maps which are weak in the corresponding intensity ($<50\;$erg$\;$cm$^{-2}\,$s$^{-1}\,$sr$^{-1}\,$\AA$^{-1}$) in the spectral raster are smoothed with a Gaussian of standard deviation of \SI[per-mode=symbol]{1}{\kilo\metre\second\tothe{-1}}. In the second step, the whole Doppler velocity map is smoothed with a Gaussian with a standard deviation of \SI[per-mode=symbol]{0.5}{\kilo\metre\second\tothe{-1}}. Contours are then defined for regions which have a blueshift stronger than \SI[per-mode=symbol]{-6}{\kilo\metre\second\tothe{-1}}. This limit was chosen by using different thresholds. A higher threshold results in the mergence of two separate events most obviously seen in the jet events. On the other hand, a lower threshold did not highlight small events and does split up events into separate contours. Increasing the threshold by \SI[per-mode=symbol]{1}{\kilo\metre\second\tothe{-1}} roughly halves the number of events, while lowering it by \SI[per-mode=symbol]{1}{\kilo\metre\second\tothe{-1}} doubles the number of events. The smoothing process influences the number of events as well. It washes out small events, which only cover single pixels. This degradation is accepted as a trade-off for smoother and closed contours.

To identify any dynamics occurring in the blueshifted regions selected, all SDO/AIA \citep{pesnell2011solar, lemen2011atmospheric} imaging data in the EUV wavebands 94, 131, 171, 193, 211, 304 and 335$\,$ \AA\ were analysed. For each feature, an event time is calculated from the central point of the blueshifted contour in the EIS raster. The SDO/AIA data are then analysed for a duration of $\pm$\SI{1}{\hour} around this event time. The data are processed with the IDL function aia\_prep and normalised for the exposure time.

To align the datasets from Hinode/EIS and SDO/AIA a cross-correlation method is used. To do this the EIS intensity maps in the Fe XII 195$\,$\AA\ emission line were used with the corresponding AIA 193$\,$\AA\ filter band images. First, an offset between the two datasets is chosen by hand to determine the initial configuration for the correlation. Both the EIS and AIA maps are then normalised and the spatial resolution of the AIA maps is reduced to that of the Hinode/EIS rasters. The final step for the correlation is to subtract the mean from each map. Then an offset for each data set can be determined. The resulting offset is applied to Hinode/EIS maps to match the SDO/AIA coordinate system.

In the last step, data from two additional instruments, Hinode/XRT and SDO/HMI, are also analysed to see whether the events have X-ray jets and if there are magnetic field changes associated with the event. Five of the events are covered by Hinode/XRT with a similar field of view as the Hinode/EIS data. Four other events are covered in XRT with high cadence full-disc data, which does not allow a proper analysis due to the low spatial resolution. Hence, those events are not further analysed in XRT. For SDO/HMI only four examples are shown since most of the datasets are too close to the limb for reliable magnetic field measurements.

\section{Categorisation of Blueshifted Events}
\label{s:Results} 

\begin{figure}
\centerline{
\includegraphics[width=0.7\linewidth]{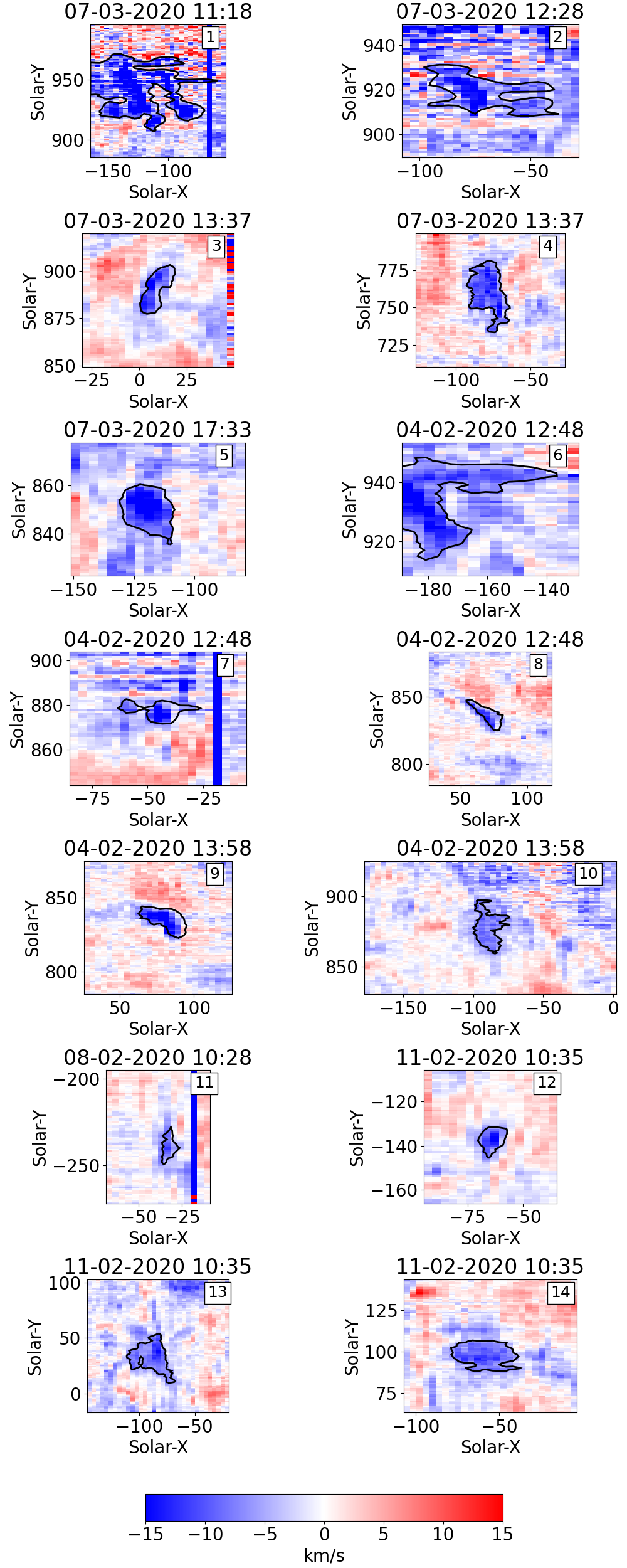}
          }
\caption{The 14 events and their \SI[per-mode=symbol]{-6}{\kilo\metre\second\tothe{-1}} blueshift contours are given with their event number, the date and time of the Hinode/EIS raster in UT in the top label. Some events have bad or no data for certain slit positions. These bad data become visible as vertical lines of strong blue- or redshifts and are ignored in the analysis.}
\label{figure:events}
\end{figure}

The Hinode/EIS contours lead to 14 blueshifted events that were analysed further. Three events are in coronal holes (CHs), five are at coronal hole boundaries and six are in the quiet Sun (QS). It is very difficult to determine whether individual upflows in the QS or CH would reach PSP, which is why it is important to get a better understanding of the physics driving each event. Most of the events are close to a pole. Those positions do not say anything about the appearance of the events but are purely based on the selection of the Hinode/EIS rasters. The events are displayed with their corresponding \SI[per-mode=symbol]{-6}{\kilo\metre\second\tothe{-1}} contours in Figure \ref{figure:events}. The exact locations and event times are given in Table \ref{table:events_details} with the sizes of the contours. The event sizes range from about \SI{100}{\sqrarcsecond} to \SI{3000}{\sqrarcsecond}. The median value is at \SI{534}{\sqrarcsecond}, which is equal to about \SI{386}{\mega\metre\squared}.

\begin{table}
\begin{tabular}{ c|c|c|c|c}
\makecell{Event \\ number} & \makecell{Centre \\ in solar-x-y} & Time &  Size  & \makecell{min, mean, max \\ velocity} \\
 1 & [-111 937] & 07-03-2020 12:17:55 & \SI{3244}{\sqrarcsecond} & [-56.6, -12.8, 12.0] km\,s$^{-1}$    \\
 2 & [-67 918] & 07-03-2020 13:18:39 & \SI{677}{\sqrarcsecond} & [-26.1, -9.7, 2.6] km\,s$^{-1}$ \\
 3 & [10 891] & 07-03-2020 14:09:00 & \SI{245}{\sqrarcsecond} & [-19.8, -10.5, -5.7] km\,s$^{-1}$ \\
 4 & [-78 758] & 07-03-2020 14:28:34 & \SI{721}{\sqrarcsecond} & [-18.2, -9.9, -4.9] km\,s$^{-1}$ \\
 5 & [-118 848] & 07-03-2020 18:32:16 & \SI{365}{\sqrarcsecond} & [-24.8, -12.6, -6.3] km\,s$^{-1}$ \\
 6 & [-170 935] & 04-02-2020 13:55:37 & \SI{699}{\sqrarcsecond} & [-22.1, -10.2, -3.3] km\,s$^{-1}$ \\
 7 & [-47 878] & 04-02-2020 13:25:13 & \SI{179}{\sqrarcsecond} & [-26.3, -10.8, 1.1] km\,s$^{-1}$ \\
 8 & [70 836] & 04-02-2020 13:02:03 & \SI{224}{\sqrarcsecond} & [-14.3, -9.8, -5.8] km\,s$^{-1}$ \\
 9 & [77 834] & 04-02-2020 14:09:27 & \SI{378}{\sqrarcsecond} & [-53.8, -15.2, -5.1] km\,s$^{-1}$ \\
 10 & [-88 879] & 04-02-2020 14:30:13 & \SI{464}{\sqrarcsecond} & [-11.2, -7.6, -5.5] km\,s$^{-1}$ \\
 11 & [-33 -238] & 08-02-2020 12:27:20 & \SI{108}{\sqrarcsecond} & [-11.9, -8.0, -5.8] km\,s$^{-1}$ \\
 12 & [-64 -138] & 11-02-2020 12:50:35 & \SI{113}{\sqrarcsecond} & [-16.5, -9.7, -5.7] km\,s$^{-1}$ \\
 13 & [-88  33] & 11-02-2020 13:08:24 & \SI{850}{\sqrarcsecond} & [-14.7, -8.3, -5.4] km\,s$^{-1}$ \\
 14 & [-58  97] & 11-02-2020 12:48:12 & \SI{539}{\sqrarcsecond} & [-13.4, -8.5, -4.3] km\,s$^{-1}$ \\
\end{tabular}
\caption{The blueshifted events are listed in an ascending number. For each event the central location of the blueshift contour in the EIS raster is given. The corresponding date and time in UT in the EIS raster of this central position is used to define the event time.}
\label{table:events_details}
\end{table}

\subsection{Analysis of All Events}

In this section, the dynamical behaviour and the surroundings of individual blueshifted events are analysed. The neighbouring features of each event are inspected. This provides an indication if there was, for example, a classic jet or a plume associated with the upflow. Clearly associated events are located below, or in the immediate vicinity of the observed location of the outflow and at the time of the EIS observation. It also allows an understanding of whether the feature is in a coronal hole, at the coronal hole boundary or in the quiet Sun. The AIA movies were made using a 12-second cadence, which revealed many small-scale dynamics during most event times. The field-of-view (FOV) was chosen in a first step to be \SI{200}{\arcsecond}$\times$\SI{200}{\arcsecond} to examine the broader surroundings. This FOV is referred to as the "vicinity" of an event throughout the paper. In a second step, a FOV slightly larger than the blueshift contour was chosen to study the behaviour within the contour. To increase the visibility of the smallest and faintest features, 30-min running difference movies were used. In addition, light curves of the average intensity within the event contours were calculated. In the last step, the location of the blueshifted sources is examined in Hinode/XRT data to look for corresponding X-ray sources.

\begin{table}
\begin{tabular}{ c|c|c|c|c|c}
Event & \makecell{Contour \\ on BP} & \makecell{BP \\ in vicinity} & \makecell{Dynamical \\ behaviour} & \makecell{XRT \\ data} & \makecell{Probable \\ source} \\
 1 & \checkmark & \checkmark & \checkmark & \checkmark & obvious jets \\
 2 & \checkmark & $\times$ & \checkmark & \checkmark & small-scale eruption \\
 3 & $\times$ & \checkmark & \checkmark & \checkmark & small-scale brightening\\
 4 & $\times$ & \checkmark & \checkmark & \checkmark & small-scale eruption \\ 
 5 & \checkmark & \checkmark & \checkmark & \checkmark & unclear  \\
 6 & \checkmark & \checkmark & \checkmark & $\times$ & bright point with jet  \\
 7 & $\times$ & \checkmark & $\times$ & $\times$ & small-scale brightening \\
 8 & $\times$ & \checkmark & \checkmark & $\times$ & small-scale eruption \\
 9 & $\times$ & \checkmark & \checkmark & $\times$ & small-scale eruption \\
 10 & $\times$ & $\times$ & \checkmark & $\times$ & small-scale brightenings \\
 11 & \checkmark & $\times$ & \checkmark & $\times$ & bright point \\
 12 & \checkmark & $\times$ & \checkmark & $\times$ & bright point \\
 13 & $\times$ & \checkmark & $\times$ & $\times$ & bright point \\
 14 & $\times$ & $\times$ & $\times$ & $\times$ & unclear \\
\end{tabular}
\caption{The blueshifted events are categorised on whether their contours overlap at least partly with a bright point (BP), whether there is a BP in the vicinity, whether there is dynamical behaviour in the contour for SDO/AIA 193\,\AA, whether there is data available in XRT. As can be seen that except for two events all others are either at least partly on a BP or have a BP in the vicinity. Furthermore, most events show dynamics in the AIA data within the contour and at the time of the event. This results in only two events, which could not be associated with a potential source at all.}
\label{table:characteristics}
\end{table}

Out of the 14 blueshifted events in Table \ref{table:characteristics}, only two could not be associated with a potential source. For all other events, probable sources could be found. This includes two with jets, four with bright points, and seven with small structures, which are mostly faint and short-lived. Jet events are blueshifted patches, where the AIA data shows at least one classical jet. In contrast to that, small-scale eruptions are events for which the image data shows a faint plasma flow, which is not a classical jet. They are sometimes seen with erupting mini-filament-like structures. Some do not show an outflow but just a faint brightening. They are referred to as small-scale brightenings. Another group of events is associated with bright points. They are seen on or next to restructuring bright points. A few events cannot be associated with any potential source, which is why they are referred to as unclear source events. However, those categories are not exclusive and some blueshift contours are caused by a combination. In the following subsections, different events for each category are presented in more detail. The events are chosen since they give a diverse overview of the different features that were seen to cause blueshifted regions.

\subsection{Jets}

\begin{figure}
\centerline{
\includegraphics[width=\linewidth]{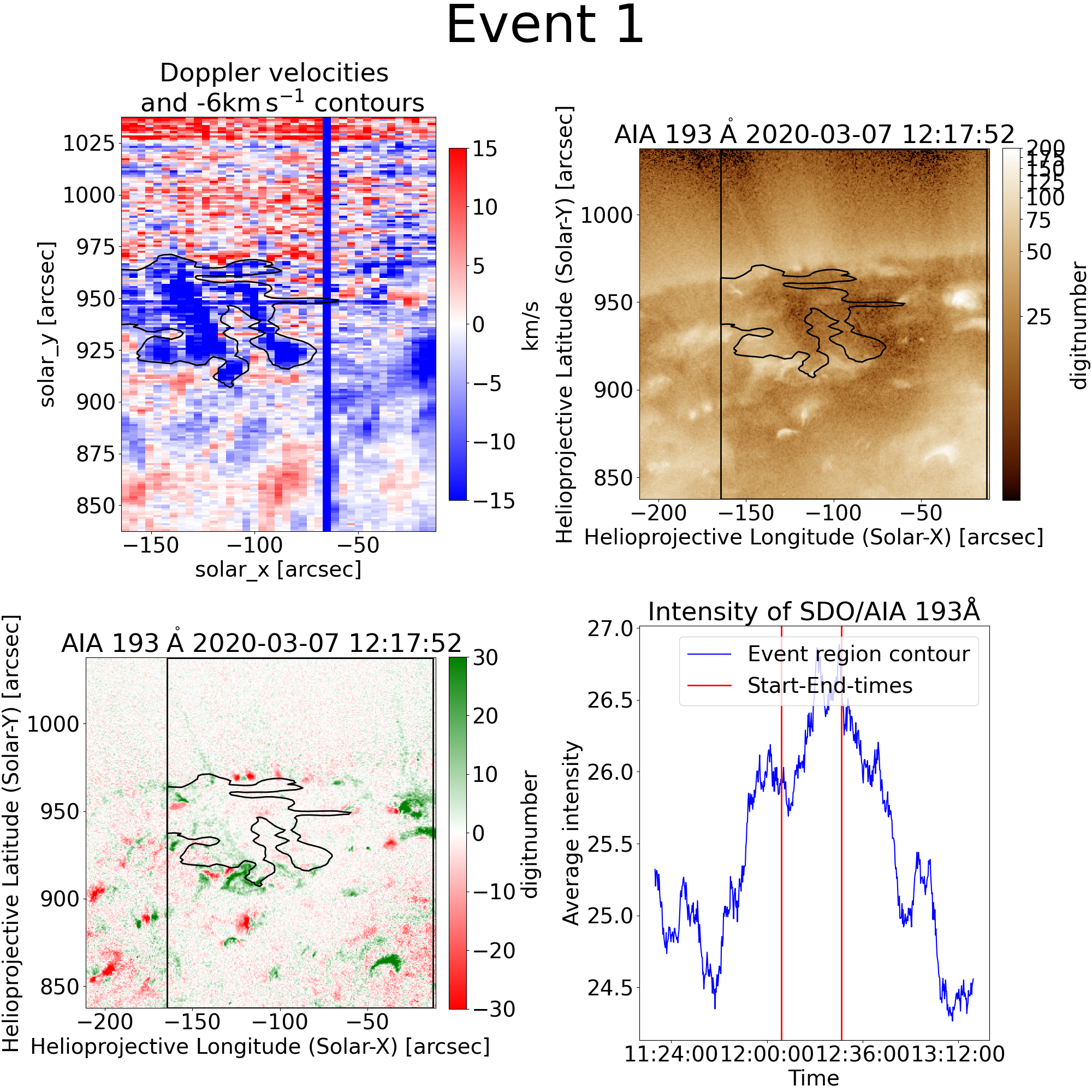}
          }
\caption{Blueshift event 1 (top-left) is a jet example. It covers a large region close to the limb. It consists of two parallel strong blueshifted patches. The corresponding AIA data (top-right) shows that it is located at a coronal hole boundary and next to a strong bright point. A jet is present at [-130, 920]. The jet is even more prominent in the difference image (bottom-left) as a brightening feature. The intensity curve of the contour (bottom-right) shows an increase over time, which peaks at the event time. The start time and end time of the blueshift contour is indicated by red lines.}
\label{figure:event0}
\end{figure}

From the first 14 blueshifted events found, only 2 were associated with clear classic jets. However, none of these events is a single classic jet, but they appear in a combination of jets or with other features, or the jet is really weak in AIA intensity images. An example of a blueshifted event likely due to a jet is event number 1, which is the largest event presented in this article with a size of \SI{3244}{\sqrarcsecond}. The contour is located directly underneath the limb. The Hinode/EIS Doppler velocities in Figure \ref{figure:event0} in the top-left plot show two elongated structures with a north-south orientation. Besides those two prominent blueshifts, smaller, and patches of weaker blueshifts lie in the eastern part of the contour. The corresponding SDO/AIA image in the top-right shows bright regions around the contour and a jet at the limb. To better visualise prominent features, a 30-minutes difference image is shown in the bottom-left. This helps to see whether bright points brighten or darken over time and to highlight short-lived features. The jet within the contour region and further jets at the limb then show a strong increase in intensity. The intensity curve of the points in the contour region in SDO/AIA in the bottom-right has a clear intensity peak. \\

When viewing this event in SDO/AIA movies in different wavelengths, at least three clear jets within the contour are visible. The first one is at [-125, 920] at 12:18UT. The second one is at [-85, 920] at 12:25UT. The third one is at [-95, 950] at 12:40UT.  The corresponding XRT data show two clear X-ray jets, which correspond to the ones seen in AIA. Besides those three jets, multiple smaller brightenings can be observed. Since the blueshift structure is not present in the prior and succeeding raster, we consider it likely that the two strong observed upflow features in Hinode/EIS are caused by those three jets, while the other regions might correspond to smaller features. A clear association between the upflows and the jets cannot be made due to the lower time and spatial resolution of Hinode/EIS in comparison to SDO/AIA and Hinode/XRT.

\subsection{Small-Scale Brightenings}

\begin{figure}
\centerline{
\includegraphics[width=\linewidth]{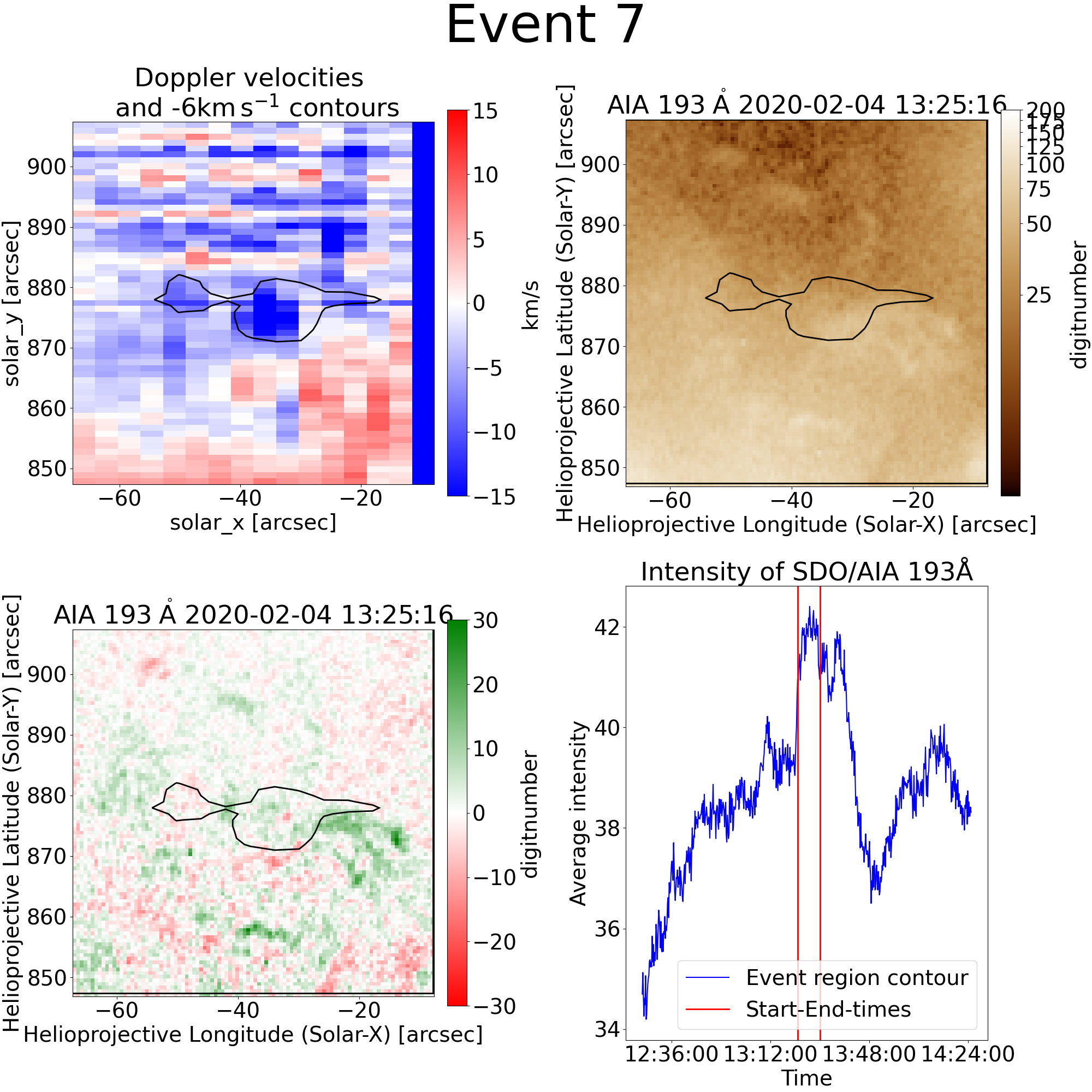}
          }
\caption{Event 7 is an example of a small-scale brightening and a rather small event. It covers only a few strongly blueshifted pixels in Hinode/EIS data (top-left). The \SI[per-mode=symbol]{-6}{\kilo\metre\second\tothe{-1}} velocity contour consists of two connected circular shapes. The corresponding AIA data (top-right) reveals that the event is located on a coronal hole boundary. A small brightening within the contour is seen in the difference image (bottom-left). The intensity curve (bottom-right) shows that a faint brightening takes place during the time the blueshifted region has been rastered.}
\label{figure:event6}
\end{figure}

Blueshifted event number 7 in Figure \ref{figure:event6} is a small feature at [-50, 875], which leads to a circular velocity contour. It is stronger in intensity than the vicinity, which shows only weak upflows or redshifted regions. The corresponding AIA image in the top-right shows that the feature is located at a coronal hole boundary where there is no prominent structure. However, the difference image in the bottom-left shows a small brightening in the contour. This coincides with the intensity curve of the pixels in the contour, which shows a clear peak. This intensity peak is illustrated through the corresponding AIA movies. The movie shows a small feature that brightens within the blueshift contour. In the corresponding 193 \AA\ movie a potential jet spire is faintly seen but harder to distinguish from the diffuse coronal haze. This event is not covered in XRT. Another example of a small-scale brightening is event number 3, which is covered in XRT. The X-ray data in this case shows a bright point which might be the source of a jet. However, a jet is also not clearly visible here. This can be either due to strong coronal haze or a perpendicular view which is close to the line of sight.

\subsection{Small-Scale Eruptions}

\begin{figure}
\centerline{
\includegraphics[width=\linewidth]{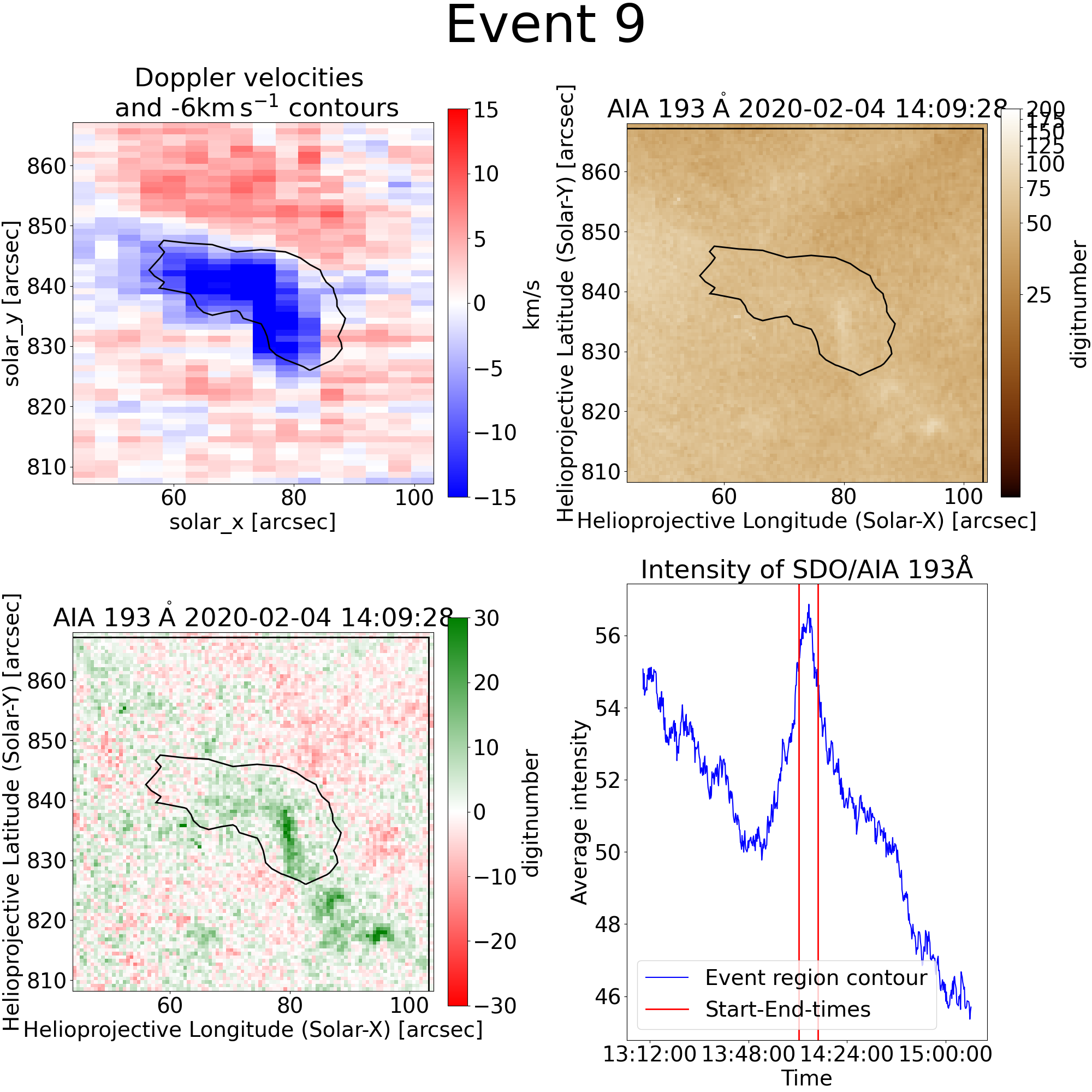}
          }
\caption{The blueshifted event 9 is an example of small-scale eruptions. In the Doppler velocity map (top-left) it shows an elongated and curved shape. It seems to continue outside of the \SI[per-mode=symbol]{-6}{\kilo\metre\second\tothe{-1}} contour. The corresponding AIA 193\,\AA\ map, shows some small bright structures directly in the contour and at its lower right. Those features can be seen much better in the difference map (bottom-left). The intensity light curve for SDO/AIA 193\,\AA\ (bottom-right) shows a clear peak at the time of the contour (red indicators). It lasts for about \SI{25}{\minute} and can be associated with the small bright points which appear, show some outflow and disappear in the corresponding movie.}
\label{figure:eruption}
\end{figure}

\begin{figure}
\centerline{
\includegraphics[width=\linewidth]{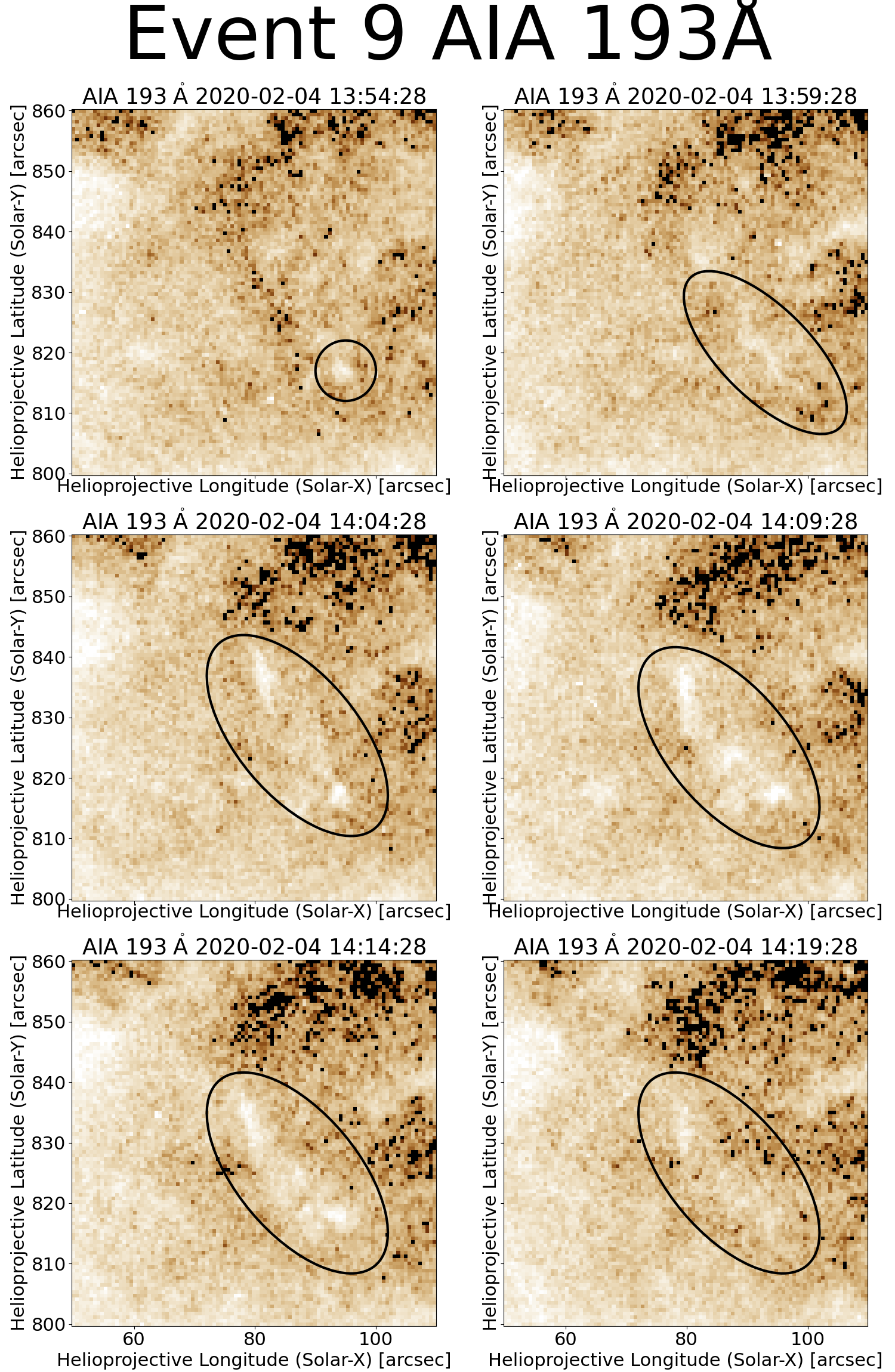}
          }
\caption{A small scale brightening followed by an outflow can be seen for event 9 in SDO/AIA 193\,\AA. It is present for about \SI{30}{\minute}.}
\label{figure:smallstructure}
\end{figure}

A large number of blueshifted regions are related to small, faint, and short-lived upflows. Those events are characterised by structures that were not clearly resolved in AIA. An example event for this group is explained in more detail and displayed in Figure \ref{figure:smallstructure}. Event 9 has an area of \SI{378}{\sqrarcsecond} and lasts for \SI{7}{\minute} in the contour. The contour has an elongated shape, which is similar to those seen for jets. The corresponding AIA movies and Figure \ref{figure:eruption} show a small structure, which appears at around 13:55 UT and disappears at around 14:25 UT. This covers the whole timeframe of the EIS contour. Only the analysis of the difference movie reveals a small bright point which causes a weak intensity flow into the event region. The corresponding light curves in EUV show peaks in the intensity, which are present in 131, 171, 193\,\AA\ and cover the same time period as the feature in the videos. In the 193\,\AA\ wavelength band the enhancement is about 13\% over about 30 minutes. This is an interesting circumstance that even though the event is weak in intensity, it lasts for an extended time. \\

This event is not covered by XRT. Examples of small-scale eruptions, which are covered in XRT, are event 2 and 4. Both events show a brightening in XRT, which could be a potential jet base bright point. However, no clear spires can be seen but only weak upflows. Hence, those events cannot clearly be labelled as classic X-ray jets. Potentially, they are either very faint jets or eruptions of different dynamics.

\subsection{Bright Points}

\begin{figure}
\centerline{
\includegraphics[width=\linewidth]{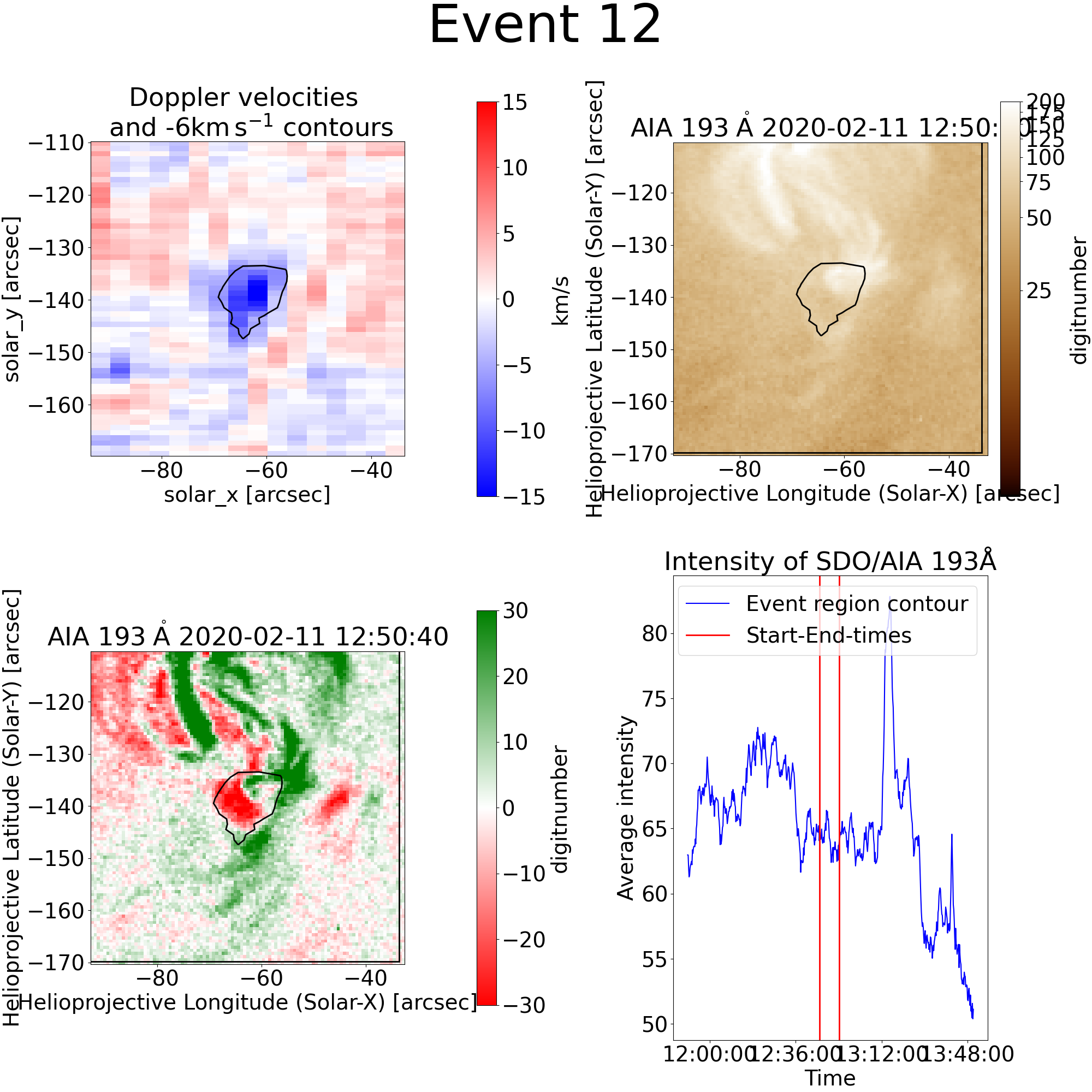}
          }
\caption{The \SI[per-mode=symbol]{-6}{\kilo\metre\second\tothe{-1}} bright point example event 12 has got a quasi-circular shape in the Doppler velocity map (top-left). When plotted onto the corresponding SDO/AIA 193\,\AA\ image (top-right), it can be seen that it partly overlaps with a bright point. The corresponding intensity light curve (bottom-left) for the SDO/AIA 193\,\AA\ data within the contour shows only small fluctuations during the exposure time of the contour (red lines). }
\label{figure:brightpoint}
\end{figure}

Many examples are connected to coronal bright points in some way. They are seen on top of bright points or next to bright points. One example is shown in more detail in this section. The presented event 12 in Figure \ref{figure:brightpoint} sits on the edge of a bright point, which is close to the equator. With a size of \SI{113}{\sqrarcsecond}, it is the second smallest upflow region. The underlying BP is much larger and consists of arch-like structures. Furthermore, it shows long and narrow extensions that reach into the contour. The horizontal extent of the contour in the EIS raster spans a time of \SI{8.5}{\minute}. Neither the movie nor the difference movie show any obvious jet or similar feature during this time. However, the bright point is dynamic and changes its structure significantly. The corresponding light curves do not show intensity peaks around the time of the blueshifted contour. There are structural changes in the bright point that are seen clearly in the difference images. 

\subsection{Unclear Sources}

\begin{figure}
\centerline{
\includegraphics[width=\linewidth]{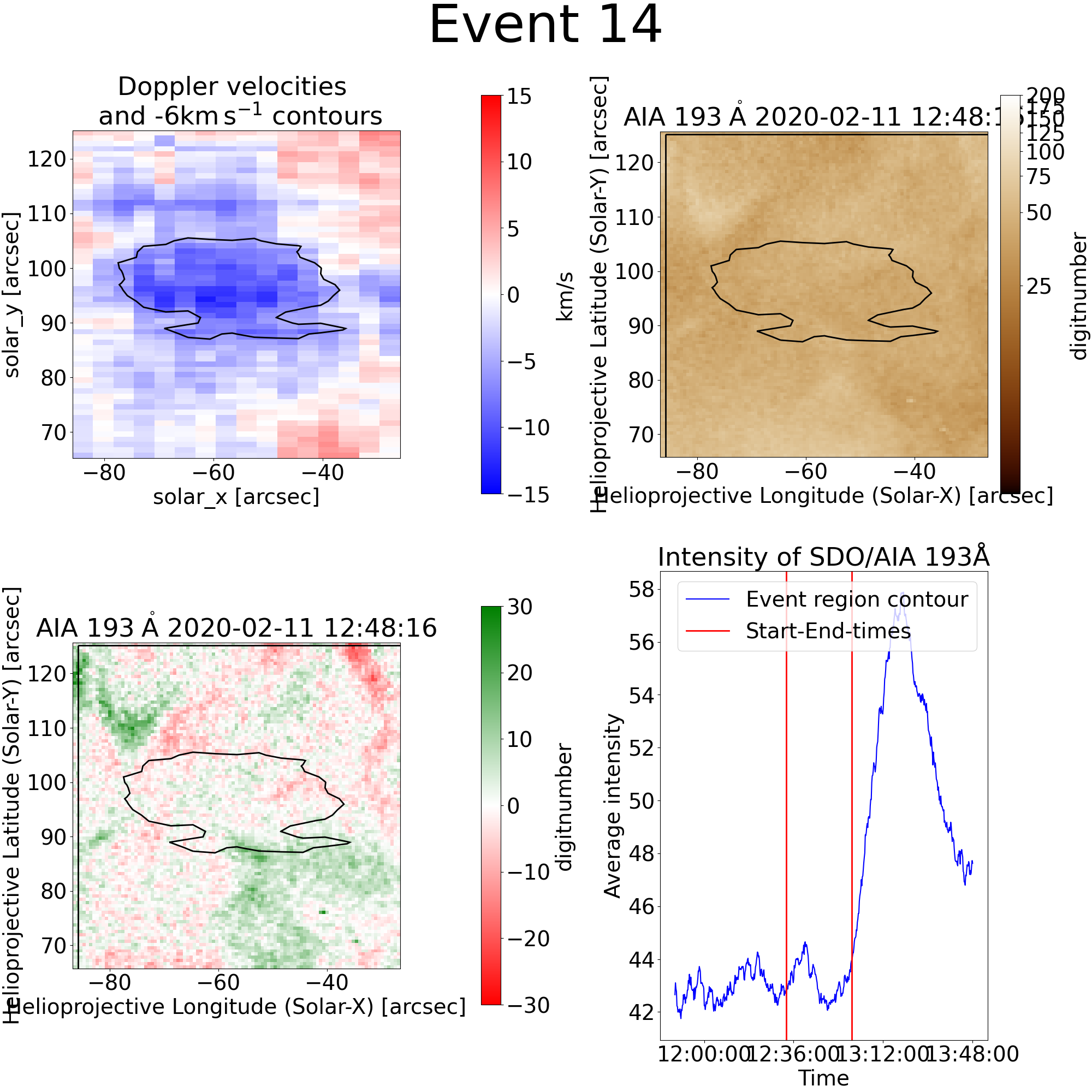}
          }
\caption{This example shows blueshifted event 14 (top-left) with no clear source, which causes a circular contour. As can be seen in SDO/AIA 193\,\AA\ (top-right) the event is located in the quiet Sun and without any bright point close by. As can be seen in the difference image (bottom-left) the region does neither brighten or darken at the event time. Also, the corresponding intensity light curve of SDO/AIA 193\,\AA\  for this event does not fluctuate strongly around the event time. Only after the event, an intensity increase can be observed}
\label{fig:nosource}
\end{figure}

In some of the events, the source remains unclear. Event 14 is an example of a no-source event shown in Figure \ref{fig:nosource}. The blueshift contour might be associated with a weak bright point and shows a circular shape. Both the AIA movies and the corresponding difference movies show only small activities before and after the event. However, no clear activity (e.g. an obvious jet), which fits in time and location of the blueshift contour, is present. The light curves in intensity do not fluctuate strongly over the whole period. An intensity increase after the event is present. This seems to correspond to a change in the vicinity of the contour and a general brightening of the region. The results do not allow to identify a clear source.

\subsection{Analysis of the Photospheric Magnetic Field Behaviour Related to Four Blueshifted Events}

To get a better understanding of the driving mechanisms for the observed events SDO/HMI data were analysed where possible. Due to the location on the disk, only the magnetic field measurements of four events could be carried out as these were close to the equator.  Those are number 11, 12, 13 and 14 in Table \ref{table:HMIcharacteristics}. 

\begin{table}
\begin{tabular}{ c|c|c }
Event & probable source & HMI feature \\
 11 & bright point & flux cancellation \\
 12 & bright point & flux cancellation \\
 13 & bright point & flux cancellation \\
 14 & unclear & none \\ 
\end{tabular}
\caption{Four blueshifted events were observed in more detail with HMI data. Three of those events are associated with bright points and one to no clear source. The three events with bright points all show flux cancellation, while the fourth event does not have any clear source in HMI.}
\label{table:HMIcharacteristics}
\end{table}

\begin{figure}
\centerline{
\includegraphics[width=\linewidth]{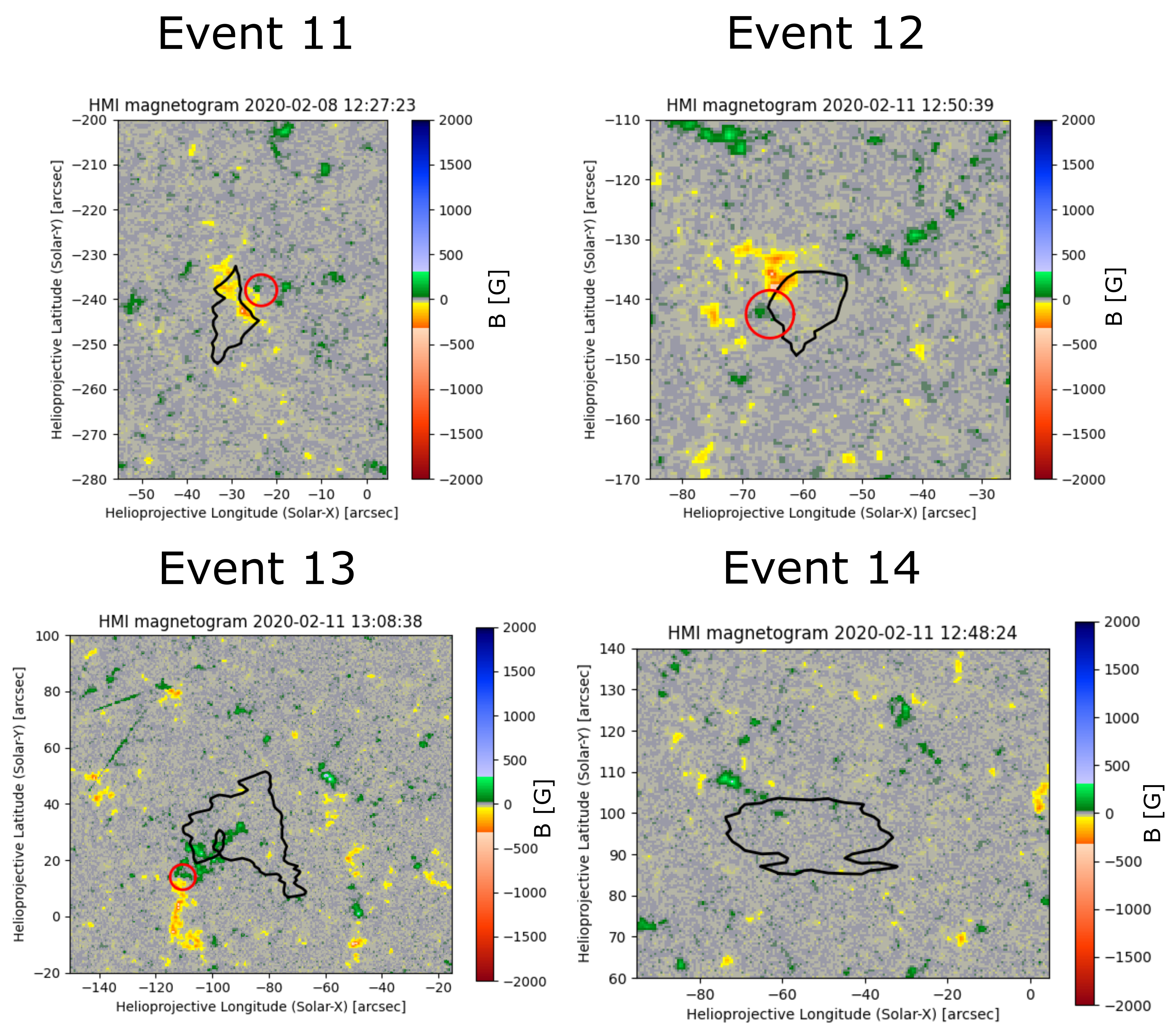}
          }
\caption{The HMI data of events 11, 12 and 13 show patches of opposite polarity close to the \SI[per-mode=symbol]{-6}{\kilo\metre\second\tothe{-1}} blueshift contour. Those fluxes cancel in the regions, which are highlighted by red circles. Event 14 in contrast to that does not show any magnetic fluxes, which can be related to the coronal event.}
\label{fig:hmievents}
\end{figure}

The magnetic configuration of event 12 in Figure \ref{fig:hmievents} shows two patches of opposite polarity. The corresponding movies show that flux cancellation takes place, where the respective regions are highlighted by a red circle. This slow magnetic restructuring of the bright point and its interaction with the surroundings could induce upflows. The upflows are located at only one end of the bright point, which also has weak structures extending into the quiet Sun that increase in intensity. This idea is supported by the observation that flux cancellation does occur at the base of jets \citep[e.g. ][]{hong2011micro, young2014coronal}. Thus there is a possibility that in this case, the cancellations led to jet-like upflows.
Events 11 and 13 which are also located on or next to a bright point, have fields of opposite polarity in Figure \ref{fig:hmievents} and flux cancellation in the corresponding HMI movies as well. This is, again, similar to event 12.
The other event is number 14 which shows no clear source in SDO/AIA data and also no significant magnetic flux changes in the vicinity of the coronal blueshifted region. This indicates that the blueshifted feature we see in the corona is either caused by photospheric changes below the resolution of HMI, or that it could be created higher in the corona.

\section{Conclusion}
\label{s:Conclusion} 

The aim of this article is to analyse regions with upflowing plasma in coronal holes and the quiet Sun. Instead of analysing the upflow of features that are prominent in intensity, such as easy-to-see jets in imaging data, the analysis started with the blueshifted plasma found using Doppler velocities in spectroscopic data. This strategy revealed blueshifted events of different sizes and velocities.

The causes for most events are diverse and mostly not related to common coronal transients such as jets. In contrast to known phenomena, many small events caused significant blueshifts which were not reported before. Potentially, the small-scale features are representations of harder-to-discern weak EUV and/or X-ray jets, which are so faint that they are not resolved properly and are blurred out by the background. All events vary in structure and appearance, but they tend to be short-lived and weak in intensity. This might be the reason why they have been missed or not commented on before.

Parker Solar Probe (PSP) has found the solar wind to be highly variable with numerous transients \citep{bale2019highly}. The source of these variations is not known.  Here we have examined the sources for 14 regions that show clear upflows in EIS Doppler data. Whether the more general upflows observed by PSP are similar to the relatively large outflow events we see here is a topic for future investigation.  

The Extreme-Ultraviolet Imager (EUI) onboard Solar Orbiter has revealed complex coronal structures on extremely small size scales down to \SI[per-mode=symbol]{400}{\kilo\metre} \citep{berghmans2021extreme}. It is of interest to see whether those small-scale features might be scaled-down versions to the BPs and possible faint jets that we find to be candidates for our 14 relatively large EIS-outflow events in this study. Also, the Solar Orbiter Polarimetric and Helioseismic Imager (PHI) instrument could look to see whether there is magnetic activity on correspondingly small scales, such as flux cancellations on a smaller scale than those that we have observed in our on-disc events here. If so, then those EUI-observed small-scale features might lead to scaled-down versions to the EIS upflows that we find in this article, and perhaps the combined upflows on the various scales can account for a portion of
the disturbances detected by PSP.

%%%%%%%%%%%%%%%%%%%%%%%%%%%%%%%%%%%%%%%%%%%%%%%%%%%%%%%%%%%%%%%%%%%%%%%%%%%
%% Appendix
%
%\appendix   

%%%%%%%%%%%%%%%%%%%%%%%%%%%%%%%%%%%%%%%%%%%%%%%%%%%%%%%%%%%%%%%%%%%%%%%%%%%
%% Acknowledgements
%
\begin{acks}
Hinode is a Japanese mission developed and launched by ISAS/JAXA, collaborating with NAOJ as a domestic partner, NASA and STFC (UK) as international partners. Scientific operation of the Hinode mission is conducted by the Hinode science team organized at ISAS/JAXA. This team mainly consists of scientists from institutes in the partner countries. Support for the post-launch operation is provided by JAXA and NAOJ (Japan), STFC (U.K.), NASA (U.S.A.), ESA, and NSC (Norway).
SDO data and images are courtesy of NASA/SDO and the AIA science team.
C.S. is grateful to SNSF for funding through project 200021\_188390. A.C.S. was supported through grants from NASA’s Heliophysics Division, and through the NASA/MSFC Hinode Project.  The work of Spanish co-authors has been supported by the Spanish Ministry for Science, Technology and Innovation through project RTI2018-096886-B-C51, and by “Centro de Excelencia Severo Ochoa” Program under grant SEV-2017-0709. D.O.S. also acknowledges financial support through the Ramón y Cajal fellowship.
\end{acks}
 \\
 \\
\noindent
\textbf{Disclosures of potential conflicts of interest:} The authors declare that they have no conflicts of interest.

%%% %%%%%%%%%%%%%%%%%%%%%%%%%%%%%%%%%%%%%%%%%%%%%%%%%%%%%%%%%%%
%% Bibliography
%
% Using BibTeX
%
 \bibliographystyle{spr-mp-sola}
 \bibliography{blueshift_bibliography}  

\begin{thebibliography}{27}
% BibTex style file: spr-mp-sola.bst (nameyear), 2015-03-09
\ifx\bisbn     \undefined \def\bisbn  #1{ISBN #1}\fi
\ifx\binits    \undefined \def\binits#1{#1}\fi
\ifx\bauthor   \undefined \def\bauthor#1{#1}\fi
\ifx\batitle   \undefined \def\batitle#1{#1}\fi
\ifx\bjtitle   \undefined \def\bjtitle#1{\textit{#1}}\fi
\ifx\bvolume   \undefined \def\bvolume#1{\textbf{#1}}\fi
\ifx\byear     \undefined \def\byear#1{#1}\fi
\ifx\bissue    \undefined \def\bissue#1{#1}\fi
\ifx\bfpage    \undefined \def\bfpage#1{#1}\fi
\ifx\blpage    \undefined \def\blpage #1{#1}\fi
\ifx\burl      \undefined \def\burl#1{\textsf{#1}}\fi
\ifx\href      \undefined \def\href#1#2{\textsf{#2}}\fi
\ifx\betal     \undefined \def\betal{\textit{et al.}}\fi
\ifx\bctitle   \undefined \def\bctitle#1{#1}\fi
\ifx\beditor   \undefined \def\beditor#1{#1}\fi
\ifx\bbtitle   \undefined \def\bbtitle#1{\textit{#1}}\fi
\ifx\bedition  \undefined \def\bedition#1{#1}\fi
\ifx\bseriesno \undefined \def\bseriesno#1{\textbf{#1}}\fi
\ifx\blocation \undefined \def\blocation#1{#1}\fi
\ifx\bsertitle \undefined \def\bsertitle#1{\textit{#1}}\fi
\ifx\bsnm      \undefined \def\bsnm#1{#1}\fi
\ifx\bsuffix   \undefined \def\bsuffix#1{#1}\fi
\ifx\bparticle \undefined \def\bparticle#1{#1}\fi
\ifx\barticle  \undefined \def\barticle#1{}\fi
\ifx\binstitute  \undefined \def\binstitute#1{#1}\fi
\ifx\bpublisher  \undefined \def\bpublisher#1{#1}\fi
\ifx\doiurl    \undefined
  \def\doiurl#1{\href{http://dx.doi.org/#1}{\textsf{DOI}}}\fi
\ifx\arxivurl  \undefined
  \def\arxivurl#1{\href{http://arxiv.org/abs/#1}{\textsf{arXiv}}}\fi
\ifx\adsurl    \undefined
  \def\adsurl#1{\href{http://adsabs.harvard.edu/abs/#1}{\textsf{ADS}}}\fi
\ifx\botherref \undefined \def\botherref#1{}\fi
\ifx\url       \undefined \def\url#1{\textsf{#1}}\fi
\ifx\bchapter  \undefined \def\bchapter#1{}\fi
\ifx\bbook     \undefined \def\bbook#1{}\fi
\ifx\bcomment  \undefined \def\bcomment#1{#1}\fi
\ifx\oauthor   \undefined \def\oauthor#1{#1}\fi
\ifx\citeauthoryear \undefined\def \citeauthoryear#1{#1}\fi
\ifx\endbibitem\undefined \def\endbibitem{}\fi
\ifx\bconflocation  \undefined \def\bconflocation#1{#1} \fi

\bibitem[\protect\citeauthoryear{Alipour and
  Safari}{2015}]{alipour2015statistical}
\begin{barticle}
\bauthor{\bsnm{Alipour}, \binits{N.}},
\bauthor{\bsnm{Safari}, \binits{H.}}:
\byear{2015},
\batitle{Statistical properties of solar coronal bright points}.
\bjtitle{Astrophysical J.}
\bvolume{807}(\bissue{2}),
\bfpage{175}.
\doiurl{10.1088/0004-637X/807/2/175}.
\end{barticle}
\endbibitem

\bibitem[\protect\citeauthoryear{Bale \textit{et~al.}}{2019}]{bale2019highly}
\begin{barticle}
\bauthor{\bsnm{Bale}, \binits{S.}},
\bauthor{\bsnm{Badman}, \binits{S.}},
\bauthor{\bsnm{Bonnell}, \binits{J.}},
\bauthor{\bsnm{Bowen}, \binits{T.}},
\bauthor{\bsnm{Burgess}, \binits{D.}},
\bauthor{\bsnm{Case}, \binits{A.}},
\bauthor{\bsnm{Cattell}, \binits{C.}},
\bauthor{\bsnm{Chandran}, \binits{B.}},
\bauthor{\bsnm{Chaston}, \binits{C.}},
\bauthor{\bsnm{Chen}, \binits{C.}}, \betal:
\byear{2019},
\batitle{Highly structured slow solar wind emerging from an equatorial coronal
  hole}.
\bjtitle{Nature}
\bvolume{576}(\bissue{7786}),
\bfpage{237}.
\doiurl{10.1038/s41586-019-1818-7}.
\end{barticle}
\endbibitem

\bibitem[\protect\citeauthoryear{Berghmans
  \textit{et~al.}}{2021}]{berghmans2021extreme}
\begin{botherref}
\oauthor{\bsnm{Berghmans}, \binits{D.}},
\oauthor{\bsnm{Auch{\`e}re}, \binits{F.}},
\oauthor{\bsnm{Long}, \binits{D.}},
\oauthor{\bsnm{Soubri{\'e}}, \binits{E.}},
\oauthor{\bsnm{Mierla}, \binits{M.}},
\oauthor{\bsnm{Zhukov}, \binits{A.}},
\oauthor{\bsnm{Sch{\"u}hle}, \binits{U.}},
\oauthor{\bsnm{Antolin}, \binits{P.}},
\oauthor{\bsnm{Harra}, \binits{L.}}, et al.:
2021,
Extreme-uv quiet sun brightenings observed by the solar orbiter/eui.
\textit{Astron. Astrophys.}
in press.
\doiurl{10.1051/0004-6361/202140380}.
\end{botherref}
\endbibitem

\bibitem[\protect\citeauthoryear{Bourouaine
  \textit{et~al.}}{2020}]{bourouaine2020turbulence}
\begin{barticle}
\bauthor{\bsnm{Bourouaine}, \binits{S.}},
\bauthor{\bsnm{Perez}, \binits{J.C.}},
\bauthor{\bsnm{Klein}, \binits{K.G.}},
\bauthor{\bsnm{Chen}, \binits{C.H.}},
\bauthor{\bsnm{Martinovi{\'c}}, \binits{M.}},
\bauthor{\bsnm{Bale}, \binits{S.D.}},
\bauthor{\bsnm{Kasper}, \binits{J.C.}},
\bauthor{\bsnm{Raouafi}, \binits{N.E.}}:
\byear{2020},
\batitle{Turbulence characteristics of switchback and nonswitchback intervals
  observed by parker solar probe}.
\bjtitle{Astrophysical J. Lett.}
\bvolume{904}(\bissue{2}),
\bfpage{L30}.
\doiurl{10.3847/2041-8213/abbd4a}.
\end{barticle}
\endbibitem

\bibitem[\protect\citeauthoryear{Culhane
  \textit{et~al.}}{2007}]{culhane2007euv}
\begin{barticle}
\bauthor{\bsnm{Culhane}, \binits{J.}},
\bauthor{\bsnm{Harra}, \binits{L.}},
\bauthor{\bsnm{James}, \binits{A.}},
\bauthor{\bsnm{Al-Janabi}, \binits{K.}},
\bauthor{\bsnm{Bradley}, \binits{L.}},
\bauthor{\bsnm{Chaudry}, \binits{R.}},
\bauthor{\bsnm{Rees}, \binits{K.}},
\bauthor{\bsnm{Tandy}, \binits{J.}},
\bauthor{\bsnm{Thomas}, \binits{P.}},
\bauthor{\bsnm{Whillock}, \binits{M.}}, \betal:
\byear{2007},
\batitle{The euv imaging spectrometer for hinode}.
\bjtitle{Sol. Phys.}
\bvolume{243}(\bissue{1}),
\bfpage{19}.
\doiurl{10.1007/s01007-007-0293-1}.
\end{barticle}
\endbibitem

\bibitem[\protect\citeauthoryear{Fisk and Schwadron}{2001}]{fisk2001behavior}
\begin{barticle}
\bauthor{\bsnm{Fisk}, \binits{L.}},
\bauthor{\bsnm{Schwadron}, \binits{N.}}:
\byear{2001},
\batitle{The behavior of the open magnetic field of the sun}.
\bjtitle{Astrophysical J.}
\bvolume{560}(\bissue{1}),
\bfpage{425}.
\doiurl{10.1086/322503}.
\end{barticle}
\endbibitem

\bibitem[\protect\citeauthoryear{Golub, Krieger, and
  Vaiana}{1975}]{golub1975observation}
\begin{barticle}
\bauthor{\bsnm{Golub}, \binits{L.}},
\bauthor{\bsnm{Krieger}, \binits{A.}},
\bauthor{\bsnm{Vaiana}, \binits{G.}}:
\byear{1975},
\batitle{Observation of a non-uniform component in the distribution of coronal
  bright points}.
\bjtitle{Sol. Phys.}
\bvolume{42}(\bissue{1}),
\bfpage{131}.
\doiurl{10.1007/BF00153290}.
\end{barticle}
\endbibitem

\bibitem[\protect\citeauthoryear{Golub \textit{et~al.}}{2008}]{golub2008x}
\begin{botherref}
\oauthor{\bsnm{Golub}, \binits{L.}},
\oauthor{\bsnm{Deluca}, \binits{E.}},
\oauthor{\bsnm{Austin}, \binits{G.}},
\oauthor{\bsnm{Bookbinder}, \binits{J.}},
\oauthor{\bsnm{Caldwell}, \binits{D.}},
\oauthor{\bsnm{Cheimets}, \binits{P.}},
\oauthor{\bsnm{Cirtain}, \binits{J.}},
\oauthor{\bsnm{Cosmo}, \binits{M.}},
\oauthor{\bsnm{Reid}, \binits{P.}},
\oauthor{\bsnm{Sette}, \binits{A.}}, et al.:
2008,
The x-ray telescope (xrt) for the hinode mission,
27.
\doiurl{10.1007/s11207-007-9058-7}.
\end{botherref}
\endbibitem

\bibitem[\protect\citeauthoryear{Harvey
  \textit{et~al.}}{1993}]{harvey1993lifetimes}
\begin{barticle}
\bauthor{\bsnm{Harvey}, \binits{K.}},
\bauthor{\bsnm{Strong}, \binits{K.}},
\bauthor{\bsnm{Nitta}, \binits{N.}},
\bauthor{\bsnm{Tsuneta}, \binits{S.}}:
\byear{1993},
\batitle{Lifetimes and distribution of coronal bright points observed with
  yohkoh}.
\bjtitle{Adv. Space Res.}
\bvolume{13}(\bissue{9}),
\bfpage{27}.
\doiurl{10.1016/0273-1177(93)90453-I}.
\end{barticle}
\endbibitem

\bibitem[\protect\citeauthoryear{Hong \textit{et~al.}}{2011}]{hong2011micro}
\begin{barticle}
\bauthor{\bsnm{Hong}, \binits{J.}},
\bauthor{\bsnm{Jiang}, \binits{Y.}},
\bauthor{\bsnm{Zheng}, \binits{R.}},
\bauthor{\bsnm{Yang}, \binits{J.}},
\bauthor{\bsnm{Bi}, \binits{Y.}},
\bauthor{\bsnm{Yang}, \binits{B.}}:
\byear{2011},
\batitle{A micro coronal mass ejection associated blowout extreme-ultraviolet
  jet}.
\bjtitle{Astrophysical J. Lett.}
\bvolume{738}(\bissue{2}),
\bfpage{L20}.
\doiurl{10.1088/2041-8205/738/2/L20}.
\end{barticle}
\endbibitem

\bibitem[\protect\citeauthoryear{Hong \textit{et~al.}}{2014}]{hong2014coronal}
\begin{barticle}
\bauthor{\bsnm{Hong}, \binits{J.}},
\bauthor{\bsnm{Jiang}, \binits{Y.}},
\bauthor{\bsnm{Yang}, \binits{J.}},
\bauthor{\bsnm{Bi}, \binits{Y.}},
\bauthor{\bsnm{Li}, \binits{H.}},
\bauthor{\bsnm{Yang}, \binits{B.}},
\bauthor{\bsnm{Yang}, \binits{D.}}:
\byear{2014},
\batitle{Coronal bright points associated with minifilament eruptions}.
\bjtitle{Astrophysical J.}
\bvolume{796}(\bissue{2}),
\bfpage{73}.
\doiurl{10.1088/0004-637X/796/2/73}.
\end{barticle}
\endbibitem

\bibitem[\protect\citeauthoryear{Kim \textit{et~al.}}{2007}]{kim2007small}
\begin{barticle}
\bauthor{\bsnm{Kim}, \binits{Y.-H.}},
\bauthor{\bsnm{Moon}, \binits{Y.-J.}},
\bauthor{\bsnm{Park}, \binits{Y.-D.}},
\bauthor{\bsnm{Sakurai}, \binits{T.}},
\bauthor{\bsnm{Chae}, \binits{J.}},
\bauthor{\bsnm{Cho}, \binits{K.S.}},
\bauthor{\bsnm{Bong}, \binits{S.-C.}}:
\byear{2007},
\batitle{Small-scale x-ray/euv jets seen in hinode xrt and trace}.
\bjtitle{Publ. Astron. Soc. JPN.}
\bvolume{59}(\bissue{sp3}),
\bfpage{S763}.
\doiurl{10.1093/pasj/59.sp3.S763}.
\end{barticle}
\endbibitem

\bibitem[\protect\citeauthoryear{Kosugi
  \textit{et~al.}}{2007}]{kosugi2007hinode}
\begin{botherref}
\oauthor{\bsnm{Kosugi}, \binits{T.}},
\oauthor{\bsnm{Matsuzaki}, \binits{K.}},
\oauthor{\bsnm{Sakao}, \binits{T.}},
\oauthor{\bsnm{Shimizu}, \binits{T.}},
\oauthor{\bsnm{Sone}, \binits{Y.}},
\oauthor{\bsnm{Tachikawa}, \binits{S.}},
\oauthor{\bsnm{Hashimoto}, \binits{T.}},
\oauthor{\bsnm{Minesugi}, \binits{K.}},
\oauthor{\bsnm{Ohnishi}, \binits{A.}},
\oauthor{\bsnm{Yamada}, \binits{T.}}, et al.:
2007,
The hinode (solar-b) mission: an overview,
5.
\doiurl{10.1007/978-0-387-88739-5_3}.
\end{botherref}
\endbibitem

\bibitem[\protect\citeauthoryear{Lemen
  \textit{et~al.}}{2011}]{lemen2011atmospheric}
\begin{botherref}
\oauthor{\bsnm{Lemen}, \binits{J.R.}},
\oauthor{\bsnm{Akin}, \binits{D.J.}},
\oauthor{\bsnm{Boerner}, \binits{P.F.}},
\oauthor{\bsnm{Chou}, \binits{C.}},
\oauthor{\bsnm{Drake}, \binits{J.F.}},
\oauthor{\bsnm{Duncan}, \binits{D.W.}},
\oauthor{\bsnm{Edwards}, \binits{C.G.}},
\oauthor{\bsnm{Friedlaender}, \binits{F.M.}},
\oauthor{\bsnm{Heyman}, \binits{G.F.}},
\oauthor{\bsnm{Hurlburt}, \binits{N.E.}}, et al.:
2011,
The atmospheric imaging assembly (aia) on the solar dynamics observatory (sdo),
17.
\doiurl{10.1007/s11207-011-9776-8}.
\end{botherref}
\endbibitem

\bibitem[\protect\citeauthoryear{Madjarska}{2011}]{madjarska2011dynamics}
\begin{barticle}
\bauthor{\bsnm{Madjarska}, \binits{M.}}:
\byear{2011},
\batitle{Dynamics and plasma properties of an x-ray jet from sumer, eis, xrt,
  and euvi a \& b simultaneous observations}.
\bjtitle{Astron. Astrophys.}
\bvolume{526},
\bfpage{A19}.
\doiurl{10.1051/0004-6361/201015269}.
\end{barticle}
\endbibitem

\bibitem[\protect\citeauthoryear{Moore
  \textit{et~al.}}{2010}]{moore2010dichotomy}
\begin{barticle}
\bauthor{\bsnm{Moore}, \binits{R.L.}},
\bauthor{\bsnm{Cirtain}, \binits{J.W.}},
\bauthor{\bsnm{Sterling}, \binits{A.C.}},
\bauthor{\bsnm{Falconer}, \binits{D.A.}}:
\byear{2010},
\batitle{Dichotomy of solar coronal jets: standard jets and blowout jets}.
\bjtitle{Astrophysical J.}
\bvolume{720}(\bissue{1}),
\bfpage{757}.
\doiurl{10.1088/0004-637X/720/1/757}.
\end{barticle}
\endbibitem

\bibitem[\protect\citeauthoryear{Nistic{\`o}
  \textit{et~al.}}{2009}]{nistico2009characteristics}
\begin{barticle}
\bauthor{\bsnm{Nistic{\`o}}, \binits{G.}},
\bauthor{\bsnm{Bothmer}, \binits{V.}},
\bauthor{\bsnm{Patsourakos}, \binits{S.}},
\bauthor{\bsnm{Zimbardo}, \binits{G.}}:
\byear{2009},
\batitle{Characteristics of euv coronal jets observed with stereo/secchi}.
\bjtitle{Sol. Phys.}
\bvolume{259}(\bissue{1-2}),
\bfpage{87}.
\doiurl{10.1007/s11207-009-9424-8}.
\end{barticle}
\endbibitem

\bibitem[\protect\citeauthoryear{Pesnell, Thompson, and
  Chamberlin}{2011}]{pesnell2011solar}
\begin{botherref}
\oauthor{\bsnm{Pesnell}, \binits{W.D.}},
\oauthor{\bsnm{Thompson}, \binits{B.J.}},
\oauthor{\bsnm{Chamberlin}, \binits{P.}}:
2011,
The solar dynamics observatory (sdo),
3.
\doiurl{10.1007/s11207-011-9841-3}.
\end{botherref}
\endbibitem

\bibitem[\protect\citeauthoryear{Raouafi and Stenborg}{2014}]{raouafi2014role}
\begin{barticle}
\bauthor{\bsnm{Raouafi}, \binits{N.-E.}},
\bauthor{\bsnm{Stenborg}, \binits{G.}}:
\byear{2014},
\batitle{Role of transients in the sustainability of solar coronal plumes}.
\bjtitle{Astrophysical J.}
\bvolume{787}(\bissue{2}),
\bfpage{118}.
\doiurl{10.1088/0004-637X/787/2/118}.
\end{barticle}
\endbibitem

\bibitem[\protect\citeauthoryear{Raouafi
  \textit{et~al.}}{2008}]{raouafi2008evidence}
\begin{barticle}
\bauthor{\bsnm{Raouafi}, \binits{N.-E.}},
\bauthor{\bsnm{Petrie}, \binits{G.}},
\bauthor{\bsnm{Norton}, \binits{A.}},
\bauthor{\bsnm{Henney}, \binits{C.}},
\bauthor{\bsnm{Solanki}, \binits{S.}}:
\byear{2008},
\batitle{Evidence for polar jets as precursors of polar plume formation}.
\bjtitle{Astrophysical J. Lett.}
\bvolume{682}(\bissue{2}),
\bfpage{L137}.
\doiurl{10.1086/591125}.
\end{barticle}
\endbibitem

\bibitem[\protect\citeauthoryear{Raouafi
  \textit{et~al.}}{2016}]{raouafi2016solar}
\begin{barticle}
\bauthor{\bsnm{Raouafi}, \binits{N.}},
\bauthor{\bsnm{Patsourakos}, \binits{S.}},
\bauthor{\bsnm{Pariat}, \binits{E.}},
\bauthor{\bsnm{Young}, \binits{P.}},
\bauthor{\bsnm{Sterling}, \binits{A.}},
\bauthor{\bsnm{Savcheva}, \binits{A.}},
\bauthor{\bsnm{Shimojo}, \binits{M.}},
\bauthor{\bsnm{Moreno-Insertis}, \binits{F.}},
\bauthor{\bsnm{DeVore}, \binits{C.}},
\bauthor{\bsnm{Archontis}, \binits{V.}}, \betal:
\byear{2016},
\batitle{Solar coronal jets: observations, theory, and modeling}.
\bjtitle{Space Sci. Rev.}
\bvolume{201}(\bissue{1-4}),
\bfpage{1}.
\doiurl{10.1007/s11214-016-0260-5}.
\end{barticle}
\endbibitem

\bibitem[\protect\citeauthoryear{Shibata
  \textit{et~al.}}{1992}]{shibata1992observations}
\begin{barticle}
\bauthor{\bsnm{Shibata}, \binits{K.}},
\bauthor{\bsnm{Ishido}, \binits{Y.}},
\bauthor{\bsnm{Acton}, \binits{L.W.}},
\bauthor{\bsnm{Strong}, \binits{K.T.}},
\bauthor{\bsnm{Hirayama}, \binits{T.}},
\bauthor{\bsnm{Uchida}, \binits{Y.}},
\bauthor{\bsnm{McAllister}, \binits{A.H.}},
\bauthor{\bsnm{Matsumoto}, \binits{R.}},
\bauthor{\bsnm{Tsuneta}, \binits{S.}},
\bauthor{\bsnm{Shimizu}, \binits{T.}}, \betal:
\byear{1992},
\batitle{Observations of x-ray jets with the yohkoh soft x-ray telescope}.
\bjtitle{Publ. Astron. Soc. JPN.}
\bvolume{44},
\bfpage{L173}.
\doiurl{10.1017/S0252921100029341}.
\end{barticle}
\endbibitem

\bibitem[\protect\citeauthoryear{Sterling
  \textit{et~al.}}{2015}]{sterling2015small}
\begin{barticle}
\bauthor{\bsnm{Sterling}, \binits{A.C.}},
\bauthor{\bsnm{Moore}, \binits{R.L.}},
\bauthor{\bsnm{Falconer}, \binits{D.A.}},
\bauthor{\bsnm{Adams}, \binits{M.}}:
\byear{2015},
\batitle{Small-scale filament eruptions as the driver of x-ray jets in solar
  coronal holes}.
\bjtitle{Nature}
\bvolume{523}(\bissue{7561}),
\bfpage{437}.
\doiurl{10.1038/nature14556}.
\end{barticle}
\endbibitem

\bibitem[\protect\citeauthoryear{Sterling
  \textit{et~al.}}{2016}]{sterling2016minifilament}
\begin{barticle}
\bauthor{\bsnm{Sterling}, \binits{A.C.}},
\bauthor{\bsnm{Moore}, \binits{R.L.}},
\bauthor{\bsnm{Falconer}, \binits{D.A.}},
\bauthor{\bsnm{Panesar}, \binits{N.K.}},
\bauthor{\bsnm{Akiyama}, \binits{S.}},
\bauthor{\bsnm{Yashiro}, \binits{S.}},
\bauthor{\bsnm{Gopalswamy}, \binits{N.}}:
\byear{2016},
\batitle{Minifilament eruptions that drive coronal jets in a solar active
  region}.
\bjtitle{Astrophysical J.}
\bvolume{821}(\bissue{2}),
\bfpage{100}.
\doiurl{10.3847/0004-637X/821/2/100}.
\end{barticle}
\endbibitem

\bibitem[\protect\citeauthoryear{Tsuneta
  \textit{et~al.}}{2008}]{tsuneta2008solar}
\begin{barticle}
\bauthor{\bsnm{Tsuneta}, \binits{S.}},
\bauthor{\bsnm{Ichimoto}, \binits{K.}},
\bauthor{\bsnm{Katsukawa}, \binits{Y.}},
\bauthor{\bsnm{Nagata}, \binits{S.}},
\bauthor{\bsnm{Otsubo}, \binits{M.}},
\bauthor{\bsnm{Shimizu}, \binits{T.}},
\bauthor{\bsnm{Suematsu}, \binits{Y.}},
\bauthor{\bsnm{Nakagiri}, \binits{M.}},
\bauthor{\bsnm{Noguchi}, \binits{M.}},
\bauthor{\bsnm{Tarbell}, \binits{T.}}, \betal:
\byear{2008},
\batitle{The solar optical telescope for the hinode mission: an overview}.
\bjtitle{Sol. Phys.}
\bvolume{249}(\bissue{2}),
\bfpage{167}.
\doiurl{10.1007/s11207-008-9174-z}.
\end{barticle}
\endbibitem

\bibitem[\protect\citeauthoryear{Young}{2015}]{young2015dark}
\begin{barticle}
\bauthor{\bsnm{Young}, \binits{P.R.}}:
\byear{2015},
\batitle{Dark jets in solar coronal holes}.
\bjtitle{Astrophysical J.}
\bvolume{801}(\bissue{2}),
\bfpage{124}.
\doiurl{10.1088/0004-637X/801/2/124}.
\end{barticle}
\endbibitem

\bibitem[\protect\citeauthoryear{Young and Muglach}{2014}]{young2014coronal}
\begin{barticle}
\bauthor{\bsnm{Young}, \binits{P.R.}},
\bauthor{\bsnm{Muglach}, \binits{K.}}:
\byear{2014},
\batitle{A coronal hole jet observed with hinode and the solar dynamics
  observatory}.
\bjtitle{Publ. Astron. Soc. JPN.}
\bvolume{66}(\bissue{SP1}).
\doiurl{10.1093/pasj/psu088}.
\end{barticle}
\endbibitem

\end{thebibliography}
%
% Without BibTeX 
% \begin{thebibliography}{}
% \bibitem[\protect\citeauthoryear{Author}{Year}]{key}
%   <bibliographical entry>
%
% \bibitem[\protect\citeauthoryear{}{}]{}
%   
%  
% \end{thebibliography}

\end{article} 
\end{document}